\documentclass[a4paper, aps, prl, twocolumn]{revtex4-1}
\usepackage{amsmath, amssymb}
\usepackage{graphicx}
\usepackage{comment}
\usepackage{hyperref}
\usepackage{soul,color}
\usepackage{braket}

\newcommand{\average}[1]{\left\langle{#1}\right\rangle}
\newcommand{\adag}{a^{\dagger}}
\newcommand{\gdn}{\Gamma_{\downarrow}}
\newcommand{\gphi}{\Gamma_{\phi}}
\newcommand{\sx}{\average{\sigma^x}}
\newcommand{\sy}{\average{\sigma^y}}
\newcommand{\sz}{\average{\sigma^z}}
\renewcommand{\aa}{\average{a a}}
\newcommand{\ada}{\average{\adag a}}

\newcommand{\xx}{C^{xx}}
\newcommand{\yy}{C^{yy}}
\newcommand{\zz}{C^{zz}}
\newcommand{\xy}{C^{xy}}

\newcommand{\abx}{C^{\alpha\beta}}
\newcommand{\ax}{C^{ax}}
\newcommand{\ay}{C^{ay}}

\newcommand{\exax}{\average{a \sigma^x}}
\newcommand{\exay}{\average{a \sigma^y}}
\newcommand{\exaz}{\average{a \sigma^z}}

\let\Re\relax \DeclareMathOperator\Re{\mathrm{Re}}%
\let\Im\relax \DeclareMathOperator\Im{\mathrm{Im}}%

\begin{document}

\title{Suppressing and Restoring the Dicke Superradiance Transition by Dephasing and Decay}

\author{Peter Kirton}
\affiliation{SUPA, School of Physics and Astronomy, University of St Andrews, St Andrews, KY16 9SS, United Kingdom}
\author{Jonathan Keeling}
\affiliation{SUPA, School of Physics and Astronomy, University of St Andrews, St Andrews, KY16 9SS, United Kingdom}
\date{\today}

\begin{abstract}
  We show that dephasing of individual atoms destroys the superradiance
  transition of the Dicke model, but that adding individual decay toward the
  spin down state can restore this transition.  To demonstrate this, we
  present a method to give an exact solution for the $N$ atom problem with
  individual dephasing which scales polynomially with $N$.  By comparing
  finite size scaling of our exact solution to a cumulant expansion, we
  confirm the destruction and restoration of the superradiance transition
  holds in the thermodynamic limit.
\end{abstract}

\maketitle

The Dicke model is a paradigmatic description of collective coupling between
matter and light~\cite{Dicke1954,Garraway2011}, the most dramatic consequence
of which is the phase transition to a superradiant
state~\cite{Hepp1973,Wang1973,Carmichael1973}.  When an ensemble of $N$
two-level atoms couple sufficiently strongly to a photon mode, the photon
ground state acquires a macroscopic, coherent, occupation.
The question of whether this phase transition actually occurs in the
true ground state of an isolated system is
subtle~\cite{Rzazewski1975,Nataf2010,Viehmann2011,Vukics2012,Bamba2014,Vukics2014,Jaako2016} due to the so-called $A^2$ term in the Hamiltonian.
However, for a driven, open system it is known the phase transition can
occur, and has been experimentally realized 
with ultra-cold atoms in optical cavities~\cite{Baumann2010, Baumann2011,
  Baden2014, Klinder2015, Kollar2016}.  Such experiments
naturally prompt questions of how loss mechanisms change the nature of the
transition~\cite{Dimer2007, Nagy2010, Keeling2010,Oeztop2012, Bhaseen2012,
  Torre2013,Piazza2013, Nagy2015,Torre2016}.

Much work on the open Dicke model has focused on photon loss processes~\cite{Dimer2007, Nagy2010, Keeling2010,Oeztop2012,Piazza2013}: this
is simple to analyze, it conserves the total spin projection so that
the dynamics is confined to a small subspace of Hilbert space, 
which can be parameterized by the total spin
vector~\cite{Emary2003,Emary2003a}.  Recently, the intriguing question of the
effect of individual dephasing and dissipation has been studied
by~\citet{Torre2016}. Using a path integral approach based
on Majorana fermionic representations of spins they found that 
the instability of the normal state survives the addition of dephasing.  
Such a result is intriguing, as previous
studies~\cite{SzymanskaThesis,Szymanska2003} using Maxwell-Bloch
equations showed that superradiance in such a system cannot
occur without electronic inversion, as in a regular
laser~\cite{Haken1970}.  This Letter aims to resolve this apparent
contradiction, providing important understanding of the relation
between superradiance and ``regular'' lasing.

As with all phase transitions, sharp features only arise in the
thermodynamic limit: for the Dicke model this corresponds to taking 
the number of spins $N \to \infty$,  while rescaling the matter-light
coupling $g$ such that $g^2 N$ remains finite.
Identifying a phase transition by means of exact numerics then
requires finite size scaling, to extrapolate whether a discontinuity arises in
the thermodynamic limit.  To use exact solutions, it is thus
necessary to solve the system for relatively large values of $N$.  In the
absence of individual dephasing, exact solution of the problem is
straightforward, as the size of Hilbert space required grows only polynomially
with system size.  In the presence of dephasing, this is not true:
the Hilbert space
grows exponentially with $N$. Thus, previous exact studies have been
limited to very small~\cite{Genway2014} numbers of spins.  It is therefore
hard to directly connect these results to the large $N$ mean-field
limit~\cite{Torre2013, Gelhausen2016, Torre2016}.

In this Letter we resolve the above issues by describing a technique, based on
the \textit{permutation symmetry} of the density operator, which allows us to
find exact numerical solutions in a time
that scales only polynominally with $N$.  
This allows us to make direct
comparison to the $1/N$ expansion arising from a cumulant expansion,
connecting finite size scaling to its asymptotic limit.  We find that
adding an infinitesimal amount of dephasing to the Dicke model destroys the
superradiant phase transition, but by also including spin relaxation the
transition is restored.  This provides an important example where the
exact nature of the dissipation included in a non-equilibrium problem can
dramatically change the steady state behavior.

Before discussing the full open system problem, let us first review the
ground state phase transition~\cite{Hepp1973,Wang1973,Carmichael1973}.
The Dicke  Hamiltonian describes the interaction of a single photon mode (creation operator $a^\dagger$) with an ensemble of $N$ spins (Pauli operators $\sigma^{x,y,z}_i$)
\begin{equation} \label{eqn:DickeHam}
	H = \omega_c a^\dagger a + \sum_i^N\omega_0 \sigma^z_i + g\sigma^x_i(a+a^\dagger).
\end{equation}
Here we have neglected the diamagnetic $A^2$ term. 
This is because we consider a driven system, where the above
Hamiltonian describes two low lying atomic levels, with the matter-light
coupling arising from adiabatic elimination of a Raman process via an
excited level~\cite{Baumann2010,Bhaseen2012, Baden2014}. 
In such a case, the value of $g$
depends on the optically active Raman transitions, while any diamagnetic
term depends on the (much smaller) bare coupling.  As has been seen
experimentally, such a driven system is thus not subject to the no-go
theorem~\cite{Rzazewski1975}, and so as the coupling strength $g$ is increased the ground state undergoes a phase transition from a normal state with $\average{a}=0$ to a superradiant state with $\average{a}\neq 0$ at a critical value of the light-matter interaction strength
$g_c^2N = {\omega_0\omega_c}/{2}$.
The superradiant state breaks the $\mathbb{Z}_2$ symmetry of the Hamiltonian where the replacement $\sigma^x_i\to-\sigma^x_i$, $a\to -a$ leaves the system unchanged.
Adding  photon losses at rate $\kappa$ results in a steady-state transition of the Liouvillian which is closely related to the equilibrium transition with a shifted critical coupling~\cite{Dimer2007},
$g_c^2N = {\omega_0}\left(\omega_c^2+{\kappa^2}/{4}\right)/{2\omega_c}$.
This kind of loss does not affect spin conservation, so the dynamics
remains within a given projection of the total spin.
Here, we also include individual spin decoherence processes,
described by the master equation
\begin{equation} \label{eqn:ME}
	\frac{d \rho}{dt} = -i[H,\rho] + \kappa \mathcal{D}[a] + \sum_i^N\gphi \mathcal{D}[\sigma^z_i] + \gdn \mathcal{D}[\sigma^-_i],
\end{equation}
where $\mathcal{D}[x]= x\rho x^\dagger - \frac12\{x^\dagger x, \rho\}$ 
is the Lindblad superoperator. 
This describes individual spin dephasing at rate $\gphi$ and losses at rate $\gdn$ as well as the coherent dynamics of the Hamiltonian and photon losses at rate $\kappa$.

To identify why the behavior with dephasing alone is not trivial, we consider
first the mean-field Maxwell-Bloch equations, which may be expected to hold at
large $N$~\cite{Haken1970}.  These give equations of motion for the expectation values of the complex photon
amplitude and the 3 real components of the spin at each site:
\begin{align}
  \label{eq:mfa}
  \partial_t \average{a} &= -\left(i\omega_c+\frac{\kappa}{2}\right)\average{a} - igN\sx \\
  \label{eq:mfb}
  \partial_t \sx &= -2\omega_0 \sy -\tilde\Gamma \sx \\
  \label{eq:mfc}
  \partial_t \sy &= 2\omega_0 \sx -4g\Re[\exaz] - \tilde\Gamma \sy  \\
  \label{eq:mfd}
  \partial_t \sz &= 4g\Re[\exay] -\Gamma_\downarrow (\sz+1)
\end{align}
where $\tilde\Gamma=2\gphi+\gdn/2$.
 The mean field equations can then be calculated from these by assuming that the second cumulants vanish and so e.g.\ $\exaz=\average{a}\average{\sigma^z}$.  Note that this
does not assume the collective spin representation can be applied.

One may see that for most choices of losses the stationary state of these equations supports both a ``normal'' solution with $\average{a}=0$ at small $g\sqrt{N}$,  and a superradiant solution with $\average{a}\neq 0$
at large $g\sqrt{N}$. i.e.\ there is a transition to a state with
coherent light in the cavity as $g\sqrt{N}$ increases.
The  exception to this is when $\gdn=0$, $\gphi\neq 0$ and $\omega_c \neq 0$ when the only solution possible is the normal state:  one can see from Eq.~(\ref{eq:mfd}) that
when $\gdn=0$, a solution with $\Re\average{a}  \neq 0$ requires
$\average{\sigma^y}=0$, however for $\gphi \neq 0$, Eq.~(\ref{eq:mfb}) then
implies $\average{\sigma^x}=0$ which in turn gives $\average{a}=0$. A purely imaginary $\average{a}$ can only satisfy Eq.~\eqref{eq:mfa} in the special case $\omega_c=0$, thus apart from this special point, there is no superradiant solution with pure dephasing.

When a phase transition occurs, it can also be found by considering the linear
stability of the normal state.  This is equivalent to the study of
susceptibility by~\cite{Torre2016}.  The normal state
always has $\average{a}_{ns}=\sx_{ns}=\sy_{ns}=0$.  When $\gdn \neq 0$, we may also specify $\sz_{ns}=-1$; for
$\gdn =0$, any value of $\sz_{ns}$ is possible.  
Performing linear stability analysis
for $\average{\sigma^x} = \average{\sigma^x}_{ns} + \delta x$ etc.\
gives the linearized equations
\begin{equation}
\label{eqn:flucmat}
\frac{d}{dt}\begin{pmatrix}
  \delta x \\ \delta y \\ \delta z 
\end{pmatrix} =
\begin{pmatrix}
  -\tilde\Gamma & -2\omega_0 & 0 \\ 2(\omega_0 + J\sz_{ns}) & -\tilde\Gamma & 0 \\ 0 & 0 & - \gdn 
\end{pmatrix}
\begin{pmatrix}
  \delta x \\ \delta y \\ \delta z 
\end{pmatrix},
\end{equation}
where  $J = 2g^2N \omega_c/(\omega_c^2 + \kappa^2/4)$ comes
from adiabatically eliminating $\average{a}$.   This matrix has unstable
eigenvalues if
\begin{equation}
\label{eqn:gcfull}
g^2N > g_c^2N = \frac{-1/\sz_{ns}}{2\omega_0\omega_c}\left(\omega_0^2+\frac{\tilde\Gamma^2}{4} \right)\left(\omega_c^2+\frac{\kappa^2}{4}\right)
\end{equation}
i.e.\ adding dissipation and dephasing simply shifts the critical
coupling, in apparent contradiction to the steady state solution when only
dephasing is present.  However, Eq.~(\ref{eqn:flucmat}) is
always  singular when $\gdn=0$ (i.e.\ not only at a phase boundary)  
since any value of $\sz_{ns}$ is a solution to the mean field equations.  Since
the critical $g\sqrt{N}$ increases with decreasing $|\sz_{ns}|$,  one may note that even if one normal state solution is unstable, 
a stable solution with smaller $|\sz_{ns}|$ always exists. 
Thus, a possible
resolution within mean field theory is that for $\gdn=0$, $\sz_{ns}$ becomes
restricted to a window approaching $0$ as $g\sqrt{N}$ increases. 
However, we can expect that beyond mean field theory, a unique steady state
always exists~\cite{Drummond1980, Lugiato1984,Rodriguez2016}, so this behavior
requires further analysis as discussed below.

To go beyond mean-field theory, we first introduce a technique which allows exact solution for much larger system sizes than is possible from naive exact diagonalization.
The master equation can be written as a sum of 
processes where each only affects a single site $i$ and the
photon mode.
From this it is clear that the problem has \textit{permutation symmetry}: swapping any pair of sites must leave the state unchanged. We may therefore gain a combinatoric reduction to the size of the Liouvillian. 

The underlying density matrix must respect the permutation symmetry described above. 
Therefore, each element of the density matrix (ignoring the photon) must obey:
\begin{multline*}
\bra{s^L_1 \ldots s^L_i \ldots s_{j}^L \ldots s_N^L}\rho\ket{s^R_1 \ldots s^R_i \ldots s_j^R \ldots s_N^R} \\
\equiv \bra{s^L_1 \ldots s^L_{j} \ldots s_{i}^L \ldots s_N^L}\rho\ket{s^R_1 \ldots s^R_{j} \ldots s_{i}^R \ldots s_N^R},
\end{multline*}
where $s^{L(R)}\in\{0,1\}$ labels the two spin states. 
The full density matrix then separates into sets of permutation symmetric elements. 
To find the time evolution of the system we only need to propagate a single representative element from each of these sets. 
Requiring this permutation symmetry applies directly to wavefunctions leads to the conservation of spin projection discussed earlier.
However, dephasing and loss mean that this restriction is not valid for the 
wavefunction, but survives for the elements of the density matrix.
Thus, we can account for the effects of individual dephasing and decay,
while keeping the size of the numerical problem manageable.

To construct the Liouvillian, $\mathcal{L}$,  first we generate a list which contains one element from each permutation symmetric set. 
To do this we combine the left and right indices for each spin so that the state at each site is described by a single number $\bra{s^L_i}\rho\ket{s^R_i} \to
\varsigma_i= s^L_i+2s^R_i$. 
We choose that the representative element for each set is the one where this list is monotonically increasing, $\varsigma_1 \le \varsigma_2 \ldots \le \varsigma_N$. 
The full list of states to consider is then the tensor product of this list with the photon states.

We then find how $\mathcal{L}$ acts on each representative element,
and project the resulting density matrix back onto the set of representative
elements, by putting the spin indices $\{ \varsigma_i \}$ into the
required  order.
We emphasize again that no approximations have been made in reducing the size of the Liouvillian, if the problem respects the symmetries described above then all density matrix elements in each class are equivalent and we only need to keep one. 
Standard differential equation solving routines can then be used to find dynamics while to find steady state properties Arnoldi iteration can be used to find the eigenvector of $\mathcal{L}$ with eigenvalue $0$.

To calculate quantities of interest from the full (but compressed) density matrix described above we need to find an efficient way to trace out many of the degrees of freedom associated with the spins.
For example, to find the reduced density matrix of a single spin and the photon we first identify elements which differ by at most a single value in the left and right spin states i.e.\ those which are diagonal in all the spins we wish to trace out.
Then we calculate the number of ways each individual element contributes to the sum,
\begin{math}
  C = {(N-1)!}/{n_\uparrow! n_\downarrow!},
\end{math}
where $n_{\uparrow(\downarrow)}$ is the number of spin up (down) particles in the state for the $N-1$ spins which are being traced out (therefore $n_\uparrow+n_\downarrow=N-1$).
From these reduced density matrices it is then straightforward to calculate, for example, expectation values of operators and Wigner distributions.

This technique is applicable to a variety of models, for example the Tavis-Cummings model including arbitrary, individual dissipation terms. Similar techniques have been employed to study spin ensembles~\cite{Chase08}, simple lasing models~\cite{Xu13} and  equilibrium properties of a model with a larger local Hilbert space~\cite{Zeb2016}.

\begin{figure}
 \includegraphics[width=0.95\columnwidth]{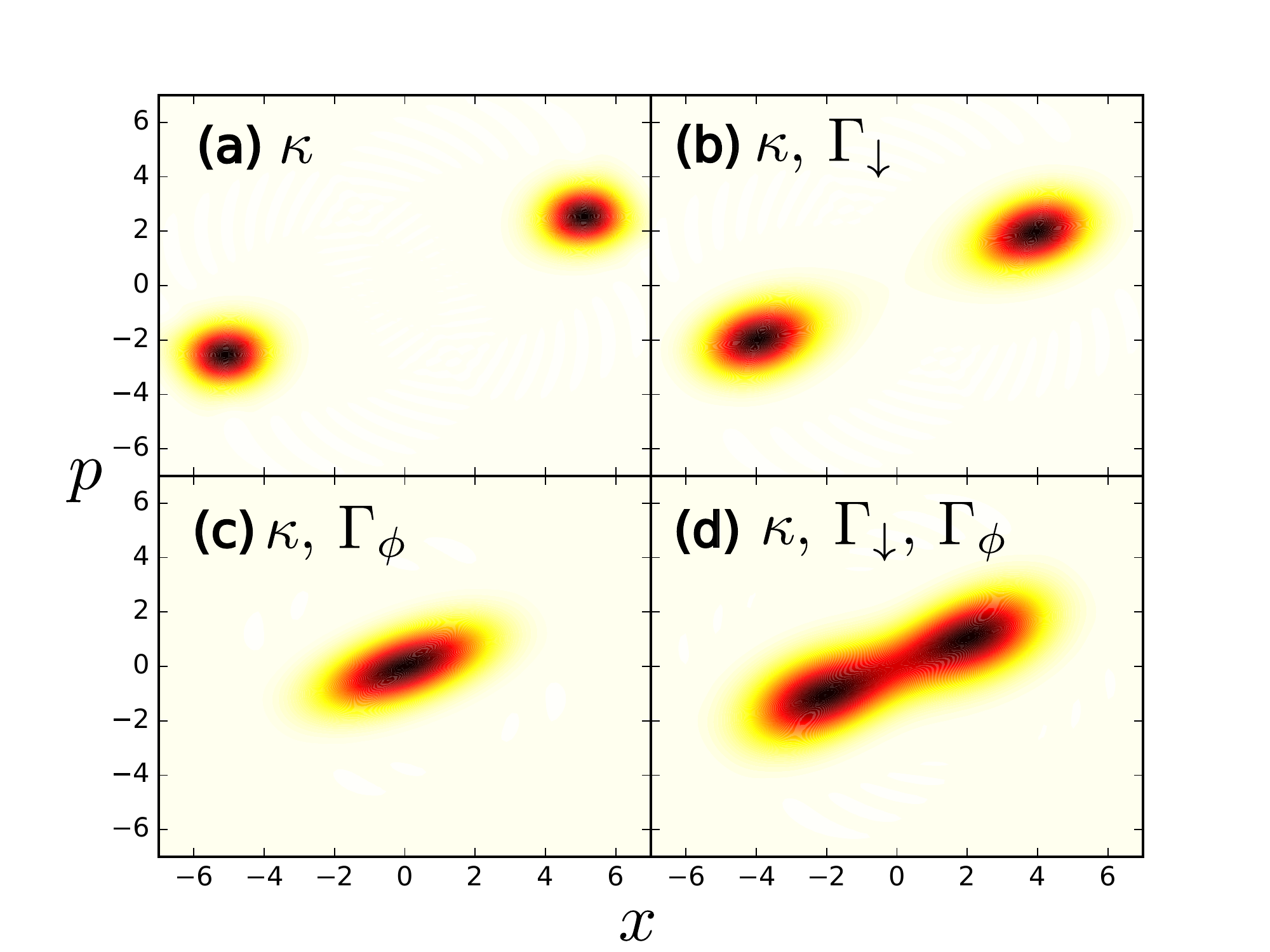}
 \caption{Wigner functions for the photon at $g\sqrt{N}=0.9$, $\omega_c=1$, $\omega_0=0.5$, $\kappa=1$ and $N=30$ for the four combinations of loss processes shown. The loss rates are $\gdn=0.2$ (b) and (d), $\gphi=0.1$ (d) and $\gphi=0.01$ (c). \label{fig:wigner}}
\end{figure}

Using our exact solution we show the steady state Wigner function of the photon for various combinations of loss processes in Fig.~\ref{fig:wigner}. 
When both $\gphi=\gdn=0$ we see two well separated peaks corresponding to the symmetry broken states in the thermodynamic limit: as expected
for a finite size system, the steady state is a mixture
of the symmetry broken states --- true symmetry breaking only
arises in the thermodynamic limit, when the tunneling time between
these solutions diverges, associated with a vanishing gap in the
spectrum of the Liouvillian~\cite{Kessler2012,Casteels2016}.
Adding spin losses as in Fig.~\ref{fig:wigner}(b) slightly reduces the amplitude of the separation and shape of these states but does not destroy the fact that two states are present.
Adding any dephasing causes the two states to coalesce destroying the possibility for a phase transition. The timescale for reaching this state diverges as the dephasing rate is reduced, but here we are interested in the asymptotic long time behavior.
Interestingly, in the presence of both losses and dephasing the indication from these results is that the transition survives.
The Wigner function has two peaks, but these are not as well separated as in the other cases. 
In what follows we will verify this behavior survives in the thermodynamic limit.

The technique we have described above is useful for calculating all of the properties of the system at relatively small values of $N$. 
To examine the nature of the phase transition at larger $N$ we will make a cumulant expansion, developing a perturbation series in powers of $1/N$ and so connecting to the small $N$ results of the exact solution.

To find the corrections of order $1/N$ we calculate the equations for the
second moments, truncating by assuming that the third cumulants vanish. Hence we split the third order moments into products of first and second order moments~\cite{Gardiner2009}.
Writing such cumulant equations allows a simple connection to approaches
widely used in laser theory~\cite{Haken1970, Haken1975}, and so provides
further conceptual clarifications.  The standard mean-field theory for the
superradiance transition is written in terms of expectations of fields that
break the $\mathbb{Z}_2$ symmetry of the Dicke model.  As such, mean-field theory is
formally incorrect for any finite size system, where quantum fluctuations and
tunneling between the solutions restores the symmetry.  
In contrast, the semiclassical rate equation frequently used in laser
theory~\cite{Haken1970,Haken1975} involves the photon
number, which does not break any symmetries of the Hamiltonian.  Moreover,
by incorporating both stimulated and spontaneous emission, one sees a sharp transition occurs only in the effective thermodynamic
limit, where spontaneous emission is weak~\cite{Haken1970}.  This
semiclassical rate equation of laser theory is analogous to solving our
model using a cumulant expansion, if one
 retains only those first- and second-order cumulants which
respect the $\mathbb{Z}_2$ symmetry of the Dicke model, and take terms such
as $\average{a}, \sx =0$. We have checked that these discarded terms make no significant changes to our results.  The phase transition is then signaled by a
discontinuity in $\average{a^\dagger a}$ which emerges when $N\to
\infty$. This discontinuity is associated with different finite size scaling in the normal
and superradiant states.

When considering equations for second moments, we may start from the
moments of the photon distribution:
\begin{align*}
  \partial_t\average{\adag a} &= -\kappa\average{\adag a} - 2gN\Im [\ax]  \nonumber \\
  \partial_t\average{aa} &= -(2i\omega_c+\kappa)  \average{aa} - 2igN\ax
\end{align*}
where we have denoted $\ax = \exax$ for brevity.
These equations involve correlations between photon and spin states, which
obey:
\begin{gather*}
  \begin{split}
  \partial_t\ax = 
  -\left(i\omega_c+\frac{\kappa}{2}+\tilde\Gamma\right)\ax - 2\omega_0 \ay 
  \\ -ig\left[\left(N-1\right)\xx +1 \right],
  \end{split}
  \\
  \begin{split}
  \partial_t\ay = 
  -\left(i\omega_c+\frac{\kappa}{2}+\tilde\Gamma\right)\ay  - 2g\sz\left(\aa+\ada\right)
  \\
    + 2\omega_0\ax - ig[(N-1)\xy-i\sz].
  \end{split}
\end{gather*}
In these equations $\abx$ means $\langle \sigma^\alpha_i \sigma^\beta_{j \neq i}\rangle$ the correlation between $\sigma^\alpha$ at one site and
$\sigma^\beta$ at another.  All such correlations are equivalent due to
permutation symmetry, while for products of operators on the same site, we
have used the standard products of Pauli matrices.  These cross correlations
obey:
\begin{align*}
  \partial_t\xx &= -4\omega_0\xy - 2\tilde\Gamma\xx, \\ 
  \partial_t\yy &= 4\omega_0\xy -2\tilde\Gamma\yy - 8g\sz\Re[\ay],
  \\
  \partial_t\zz &= 8g\sz\Re[\ay] -2\gdn(\zz +\sz),
  \\
  \partial_t\xy &= 2\omega_0(\xx-\yy)-2\tilde\Gamma\xy -
  4g\sz\Re[\ax].  
\end{align*}
These then allow us to cross from the small $N$ limit --- accessible by exact numerics --- to the large $N$ limit required to make concrete statements about phase transitions.

\begin{figure}
 \includegraphics[width=0.95\columnwidth]{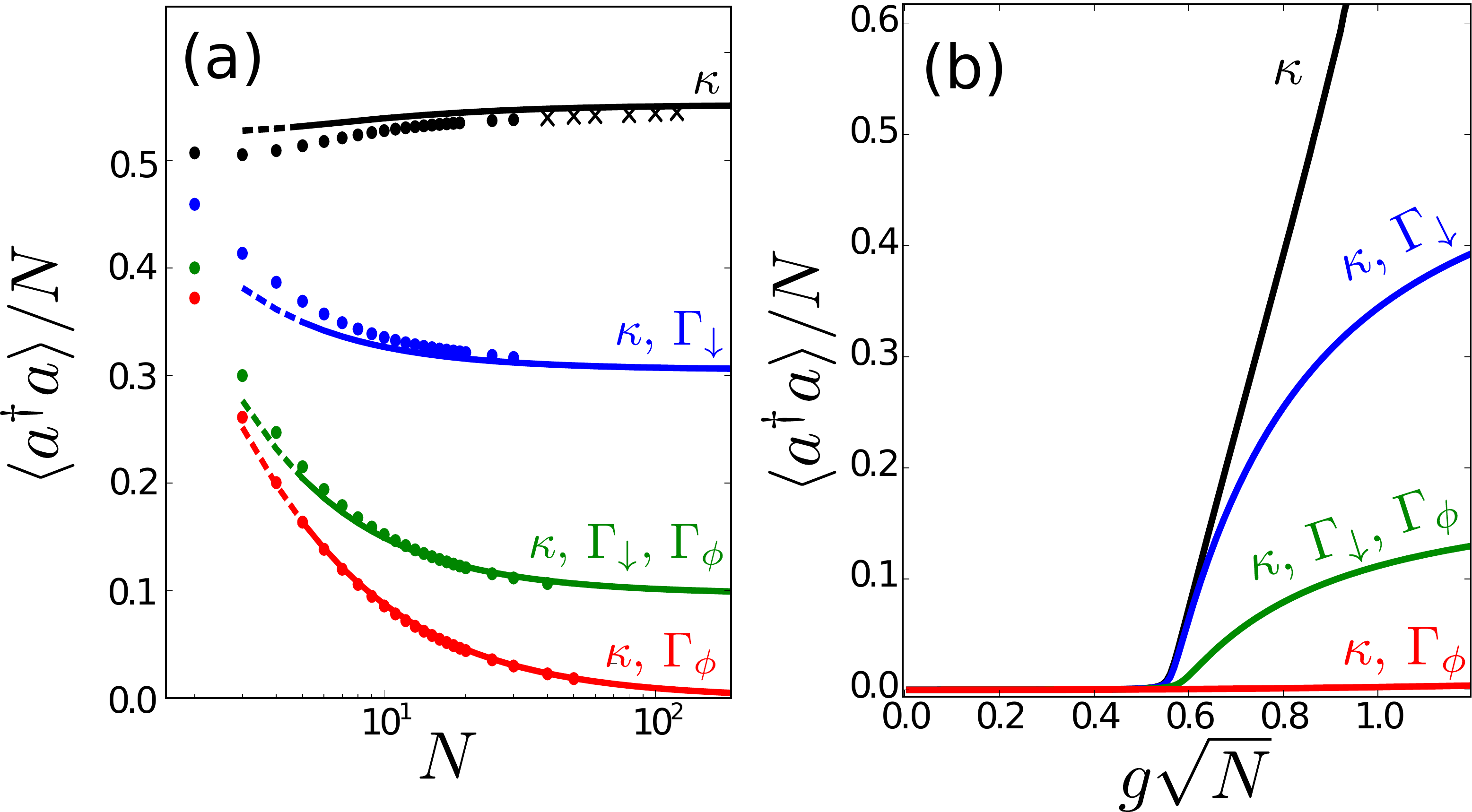}
 \caption{(a) Steady state photon number vs $N$. The labels show which loss
   processes are present. The lines correspond, from top to bottom, to
   $\gdn=\gphi=0$ (black), $\gdn=0.2$, $\gphi=0$ (blue), $\gdn=0.2$,
   $\gphi=0.1$ (green) and $\gdn=0$, $\gphi=0.01$ (red). The solid lines are
   the cumulant expansion and the dots show the numerically exact
   solution. Other parameters are: $g\sqrt{N}=0.9$, $\omega_c=1$,
   $\omega_0=0.5$, $\kappa=1$. For the $\gdn=\gphi=0$
   case, the crosses at large $N$ values result from averaging $5000$ quantum trajectories, except for $N=120$ and $N=150$ where $1000$ and $500$ trajectories were used respectively. (b) Photon number vs $g\sqrt{N}$ from the second order cumulant expansion. The lines correspond to the same loss processes and parameters as in (a) with $N=500$.\label{fig:phot_vs_N}
   }
\end{figure}

In Fig.~\ref{fig:phot_vs_N}(a) we compare the results of the exact solution to
those of the second order cumulant expansion, for the steady state photon
number as a function of $N$.  At large $N$ both approaches match and we can
be confident that the cumulant expansion gives the correct steady state
behavior.  For the case $\gdn=\gphi=0$, the exact solution can make use of
the fixed spin projection dynamics, allowing exact results to larger $N$.
For the largest values of $N$ we use a quantum trajectories
approach~\cite{carmichael2009, Johansson2013} to find the steady state.  For the other cases,
the value of $N$ achievable is limited by the size of the photon Hilbert space
needed: this size differs depending on the dephasing processes: for the case
with $\gphi=0, \gdn \neq 0$ we require photon numbers up to $35$ restricting us
to $N=30$.

By considering the finite size scaling of $\ada/N$, we see a clear
distinction that for dephasing alone, this ratio vanishes as $N \to \infty$,
but remains finite for all other cases.  This indicates that the superradiant
phase is not present for the pure dephasing case.  The value of light-matter
coupling chosen here is well above that expected for the transition
which, without spin decoherence, would be at $g_c\sqrt{N}\simeq 0.56$.
Similar calculations below this threshold show that the rescaled
photon number goes to 0 in all four cases.  We also note that the
convergence of these results is better in the
presence of dephasing.

Figure~\ref{fig:phot_vs_N}(b) shows the rescaled steady state photon number as a function of $g\sqrt{N}$ for the four different scenarios calculated from the cumulant expansion.
We clearly see the emergence of non-analytic behavior in all cases except for when only dephasing is present.
The discontinuities occur at exactly the location predicted by Eq.~\eqref{eqn:gcfull}. 
While the addition of loss processes does reduce the steady state photon number there is still a superradiant phase except in the pure dephasing model. 
As can be seen from the mean-field expressions,
the critical exponent $\beta$ defined by $\ada=(g-g_c)^{\beta/2}$ 
takes the value $\beta=2$ in each of these cases where there is a transition,
in contrast to the more complex behavior that can occur with non-Markovian loss~\cite{Nagy2015,Lang16}.

In conclusion we have explored the fate of the Dicke transition in the presence of losses and dephasing. 
We find that while adding any dephasing to the model kills the steady state phase transition, it survives the presence of spin losses. While, as found in Ref.~\cite{Torre2016},  there are values of $\sz$ for which the normal state is unstable, we have demonstrated that these do not lead to a superradiant steady state, but instead to another normal state.
Surprisingly, we find that spin losses are able to stabilize the transition to dephasing. 
To study this model we made use of exact solutions and cumulant expansions
for the finite sized system.  The cumulant expansion
makes connections to standard laser theory by showing how the photon number discontinuity
arises in the thermodynamic limit.
Our exact solution uses a new numerical technique, which scales polynomially with the number of spins, to exactly describe the system's density matrix. 
This technique can be used to study many other models with the same permutation symmetry as those described here.

\begin{acknowledgments}
  We acknowledge useful discussions with M.\ Hartmann and E.\ Dalla Torre.
  P.G.K.\ acknowledges support from EPSRC (EP/M010910/1).  J.K.\ acknowledges support
  from EPSRC program “TOPNES” (EP/I031014/1). The research data supporting this publication can be accessed at \url{http://dx.doi.org/10.17630/001a842d-94cd-4b72-9465-d1e67d43f257}
  The code for the exact solutions can be found at \url{https://dx.doi.org/10.5281/zenodo.376621}
\end{acknowledgments}


%

\end{document}